\documentclass[twocolumn,showpacs,preprintnumbers,amsmath,amssymb,floatfix,nofootinbib]{revtex4}

\usepackage{graphicx}% Include figure files
\usepackage{dcolumn}% Align table columns on decimal point
\usepackage{bm}% bold math

\begin{document}

\def\ba{\begin{eqnarray}}
\def\ea{\end{eqnarray}}
\def\be{\begin{equation}}
\def\ee{\end{equation}}
\def\({\left(}
\def\){\right)}
\def\[{\left[}
\def\]{\right]}
\def\lagrange {{\cal L}}
\def\del {\nabla}
\def\d {\partial}
\def\Tr{{\rm Tr}}
\def\half{{1\over 2}}
\def\fourth{{1\over 8}}
\def\bibi{\bibitem}
\def\S{{\cal S}}
\def\H{{\cal H}}
\def\K{{\cal K}}
\def\xx{\mbox{\boldmath $x$}}
\def\jmin{{${\cal J}^-$\,}}
\def\jmine{{{\cal J}^-}}
\def\ourmod{{${\cal M}$}}
\newcommand{\phpr} {\phi'}
\newcommand{\gam}{\gamma_{ij}}
\newcommand{\sqgam}{\sqrt{\gamma}}
\newcommand{\delk}{\Delta+3{\cal K}}
\newcommand{\dph}{\delta\phi}
\newcommand{\om} {\Omega}
\newcommand{\dom}{\delta^{(3)}\left(\Omega\right)}
\newcommand{\rar}{\rightarrow}
\newcommand{\Rar}{\Rightarrow}
\newcommand{\labeq}[1] {\label{eq:#1}}
\newcommand{\eqn}[1] {(\ref{eq:#1})}
\newcommand{\labfig}[1] {\label{fig:#1}}
\newcommand{\fig}[1] {\ref{fig:#1}}
\newcommand{\sgn}{\,{\mathrm{sgn}}}

\def\gsim{ \lower .75ex \hbox{$\sim$} \llap{\raise .27ex \hbox{$>$}} }
\def\lsim{ \lower .75ex \hbox{$\sim$} \llap{\raise .27ex \hbox{$<$}} }
\newcommand\bigdot[1] {\stackrel{\mbox{{\huge .}}}{#1}}
\newcommand\bigddot[1] {\stackrel{\mbox{{\huge ..}}}{#1}}

%\preprint{APS/123-QED}

\title{Inflation without a beginning: a null boundary proposal}

\author{Anthony Aguirre}
\affiliation{School of Natural Sciences, Institute for Advanced Study
Princeton, New Jersey 08540, USA}
\email{aguirre@ias.edu}
\author{Steven Gratton}
\affiliation{Joseph Henry Laboratories, Princeton University, Princeton, New Jersey
08544, USA}
\email{sgratton@princeton.edu}

\date{\today}

\begin{abstract}
We develop our recent suggestion that inflation may be made past
eternal, so that there is no initial cosmological singularity or
``beginning of time''.  Inflation with multiple vacua generically
approaches a steady-state statistical distribution of regions at these
vacua, and our model follows directly from making this distribution
hold at all times. We find that this corresponds (at the
semi-classical level) to particularly simple cosmological boundary
conditions on an infinite null surface near which the spacetime looks
de Sitter.  The model admits an interesting arrow of time that is
well-defined and consistent for all physical observers that can
communicate, even while the statistical description of the entire
universe admits a symmetry that includes time-reversal. Our model
suggests, but does not require, the identification of antipodal points
on the manifold.  The resulting ``elliptic'' de Sitter spacetime has
interesting classical and quantum properties. The proposal may be
generalized to other inflationary potentials, or to boundary
conditions that give semi-eternal but non-singular cosmologies.
\end{abstract}

\pacs{98.80.Cq, 04.20.Gz, 04.62.+v}

\maketitle

\section{Introduction}

	The ancient philosophical question of whether the universe is
finite or infinite in time, and whether time ``had a beginning'',
entered the domain of scientific study during the 20th century
with the development of generally relativistic cosmologies in which
the universe is homogeneous and expanding, as observed on the largest
accessible scales.

	Chief among these cosmologies, and representing opposite
answers to the question of the universe's beginning, were the
classical Big-Bang and Steady-State. As FRW models, both are based on
some form of the ``Cosmological Principle'' (CP) that the large-scale
statistical properties of the universe admit spatial translational and
rotational symmetries.  The models differ greatly, however, in their
time evolution. In the Big-Bang, the properties of the universe evolve
in a finite time from a dense, singular initial state.  In contrast,
the Steady-State universe is said to obey the ``Perfect Cosmological
Principle'' (PCP) in that it admits, in additional to spatial
translational and rotational symmetries, a time-translation symmetry.
Since all times are equivalent, there can be no ``beginning of time'',
and the universe is infinite in duration.

	Unlike their philosophical predecessors, the Big-Bang and
Steady-State model were observationally distinguishable, and
astronomical evidence eventually turned nearly all cosmologists away
from the Steady-State.  Moreover, theorems proven within General
Relativity showed that the classical singularity of the Big-Bang
cosmology was robust and could not be avoided by relaxing simplifying
assumptions such as that of homogeneity.  Thus the idea of a
temporally finite universe with a singular initial epoch came to
dominate cosmology.\footnote{There have been a number of proposals for
avoiding a beginning of time, but generally these involve either
continuing through the cosmological singularity by invoking quantum
gravitational effects, or modifying GR at the classical level.  Our
approach aims to develop a non-singular cosmology without appeal to
either possibility.}  Attention has since focused on how this
primordial singularity (where some presently unknown theory of quantum
gravity presumably applies) could give rise to a classical ``initial''
state that could evolve into the observed universe.

	The required ``initial'' classical state, however, seemed
rather special: the universe had to have been extremely flat, and
statistically homogeneous (i.e. obey the CP) on scales larger than the
horizon size.  The theory of inflation was devised and widely accepted
as a solution to this problem of a special initial state: given
inflation, a flat, homogeneous universe (with the necessary Gaussian
scale-invariant density fluctuations) is an attractor.  That is,
within some inflating region of fixed, finite physical size, the CP
holds more and more precisely with time.  What is perhaps more
surprising and less widely appreciated, however, is that in generic
inflation models the universe also comes to obey, with ever-greater
precision, the {\em Perfect} Cosmological Principle.  This occurs
because inflation is generically ``semi-eternal'': rather than ending
globally at some time, inflation always continues in some regions, and
the universe globally approaches a quasi-steady-state distribution of
inflating and thermalized regions, the statistical description of
which becomes asymptotically independent of
time~\cite{qssinfl}.

	Since inflation generically {\em approaches} a steady-state,
it seems physically reasonable to ask whether the universe can simply
{\em be} in an inflationary steady-state, thus avoiding a cosmological
singularity or ``beginning of time''.  Indeed, the possibility of
truly eternal inflation was raised soon after inflation's invention,
but no satisfactory model was immediately devised~\cite{etin}, and in
subsequent years several theorems were formulated proving the geodesic
incompleteness of models globally satisfying conditions seeming
necessary for eternal inflation~\cite{singth,bgv}.  These theorems
suggested that inflationary cosmologies necessarily contain
singularities, but the exact nature of the implied singularities was
obscure.

	In a recent paper, we constructed a counter-example to these
theorems by providing a model for geodesically complete truly eternal
inflation~\cite{ssei}.  There, we analyzed the classical Steady-State
model in detail, then extended our analysis to inflation.  Here, we
develop the model from a different standpoint, focusing on an eternal
inflation in a double-well inflaton potential, and on the
corresponding cosmological boundary conditions.
Section~\ref{sec-prop} motivates and develops our model, and describes
its general features. Various aspects of the model are developed in
subsequent sections: Section~\ref{sec-aot} discusses the arrow of time
in our model, and elucidates the failure of the singularity theorems
to forbid our construction; Sec.~\ref{sec-bcs} discusses the
cosmological boundary conditions, which are specified on a null
surface; Sec.~\ref{sec-elliptical} discusses the relation of our model
to the ``antipodally identified'' or ``elliptic'' interpretation of de
Sitter spacetime that it suggests, and briefly discusses quantum field
theory in elliptic de Sitter; Sec.~\ref{sec-general} discusses
generalizations and extensions of our model. We summarize and conclude
in Sec.~\ref{sec-conc}.

\section{The proposal}
\label{sec-prop}

In this section we develop an eternally inflating cosmology based on a
double-well inflaton potential $U(\phi)$ with minima at $\phi_t$ and
$\phi_f$, where $U(\phi_f) > U(\phi_t) \ge 0$.  This sort of potential
is posited in ``old'' inflation~\cite{guth} or ``open''
inflation~\cite{bgt,openinf}. We will first review semi-eternal
double-well inflation, then extend this to eternal inflation, then
analyze and address the geodesic completeness of the model.

\subsection{Semi-eternal ``double-well'' inflation}

A semi-eternally inflating cosmology naturally arises from a generic
double-well potential~\cite{vil92,guthwein}.  Consider some large
comoving region in which, at an initial time $t=t_0$, the
energy density is dominated by the inflaton $\phi$, with $\phi\simeq\phi_f$
and $\partial_\mu\phi\simeq 0$. Assume the spacetime to locally
resemble de Sitter (``dS'') space, with (for convenience) nearly-flat
spatial sections, i.e. with a metric~\cite{dsreview} approximated by
\begin{equation}
ds^2=-dt^2 + e^{2Ht} \(dx^2+dy^2+dz^2\).
\labeq{flatds}
\end{equation}

In this background, bubbles of true vacuum $\phi_t$ nucleate at a
fixed rate $\lambda$ per unit physical 4-volume that depends upon the
potential $U(\phi)$\cite{coldel}.  The interior of each bubble looks
like an open FRW cosmology to observers inside it. For a suitably
designed $U(\phi)$ (as in open inflation\cite{bgt,openinf}),
there can be a slow-roll inflation epoch inside the bubble so that the
FRW regions are nearly flat and homogeneous, and have scale-invariant
density perturbations. One such region could therefore in principle
represent our observable cosmological surroundings.

At any time $t$, we can derive the distribution of bubbles and
inflating region within our comoving volume, with the aim of showing
that the distribution approaches a steady-state. To avoid
complications resulting from bubble collisions and the ambiguities in
connecting the time-slicings within and outside bubbles, we
concentrate on the statistics describing the inflating region outside
of the bubbles.  This region is necessarily unaffected by
the bubbles' presence because they expand at the speed of light: both
its global and local properties depend only on its initial state at
$t_0$. But we may describe three effects of the bubble encroachment
upon it.

First, let us consider the inflating region left at time $t$ by
bubbles forming since $t_0$.  This region must consist entirely of
points each of which does {\em not} have a nucleation event in its
past light-cone (PLC) going back to $t_0$. Denoting the volume of this
PLC by $Q(t,t_0)$, it can then be shown that such points comprise a
volume fraction
\begin{equation}
f_{\rm inf} = \exp [-\lambda Q] \simeq \exp\left[\frac{-4\pi\lambda(t-t_0)}{3H^3}\right] 
\labeq{finfl}
\end{equation}
for $(t-t_0) \gg H^{-1}$\cite{vil92,guthwein}. 
Although $f_{\rm inf}\rightarrow 0$ for large $t-t_0$, the spacetime
is said to be eternally inflating because for small $\lambda$ the
physical inflating volume within our comoving region nonetheless
increases exponentially with time:
\begin{equation}
V_{\rm inf} \propto f_{\rm inf}\exp(3Ht)\sim \exp(DHt)
\labeq{vinf}
\end{equation}
for any fixed $t_0$, with $D\equiv{3-{4\pi\lambda/3H^4}}$.

Second, one can show~\cite{vil92} that at fixed $t$ the distribution
of inflating regions about any inflating point is described by a
fractal of dimension $D$ (that is, the inflating volume $V_{\rm inf}
\propto r^3f_{\rm inf}(r) \propto r^D$) up to a scale of order
$r_B(t,t_0)$, where 
\begin{equation}
r_B(t,t_0)=H^{-1}\[e^{H(t-t_0)}-1\]
\labeq{bubbrad}
\end{equation}
is the physical radius at $t$ of a bubble nucleated at $t_0$.

Third, we may calculate, for a given point in the inflating region at
time, the number per unit time $N(r,t)$ of incoming bubbles of
physical radius $r$. This is
\begin{equation}
N(r,t)={4\pi\lambda r^2\over(1+Hr)^4}
\labeq{bubbrate}
\end{equation}
for $r<r_B(t,t_0)$ and zero for $r > r_B$.

An observer within a bubble can never leave, but will eventually be
encountered by an encroaching bubble wall after a typical time
$\tau_{\rm coll}$, where $\tau_{\rm coll}^{-1}$ is related to the
$r$-integral of Eq.~\eqn{bubbrate} by some transformation between the
bubble observer's proper time $\tau$ and cosmic time $t$.
Since this rate depends on $t-t_0$, a patient and very sturdy observer
could in principle discover the global time at which it formed by
counting the frequency of incoming bubbles.

Now, as $t \rightarrow \infty$, four things occur.  First, the
nearly-flat spatial sections approach perfect flatness.  Second, the
incoming bubble rate $N(r,t)$ becomes homogeneous and independent of
time on arbitrarily large physical scales.  To see this, imagine that
the rate is inhomogeneous at early times because bubble nucleation
starts at different times in different regions.  But since the impact
rate depends on the initial time only for bubbles of radius greater
than $r_B(t,t_0)\sim\exp[H(t-t_0)]$, it is then homogeneous and
independent of time on arbitrarily large scales as
$t\rightarrow\infty$.  Third, and for essentially the same reason, the
distribution of inflating region around any given inflating region
also becomes homogeneous and independent of time on arbitrarily large
scales.  Fourth, observers within bubbles lose the information about
the ``global time'' at which they exist (see Eq.~\eqn{bubbrate}), and
all bubbles become equivalent. Thus the physical description of the
universe, relative to any fixed length scale such as $H^{-1}$,
satisfies the Perfect Cosmological Principle arbitrarily well as
$t\rightarrow \infty$.

\begin{figure}
\includegraphics[width=8.5cm]{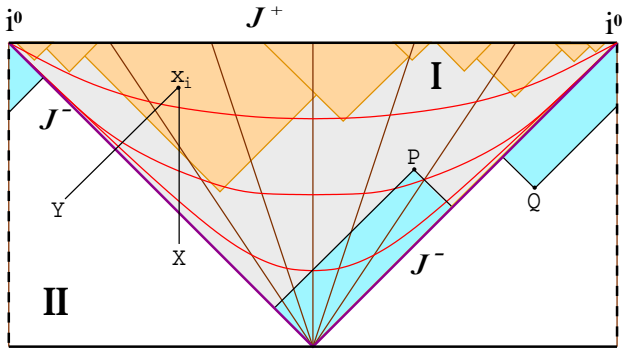}
\caption{Conformal diagram for de Sitter (dS) spacetime.  Each point
represents one point in 1+1 D dS, or half of a 2-sphere in 3+1 D dS.
The left and right (dotted) edges are identified.  The shaded region
(region I) is covered by coordinates with flat spatial sections
(spacelike lines) with spacelike infinity at $i_0$; the straight,
timelike lines represent comoving geodesics. The null surface \jmin
represents $t\rightarrow -\infty$.  True-vacuum ``test bubbles''
(not disturbing the background spacetime) are darkly shaded and
open toward future timelike infinity ${\cal J}^+$.  Also shown are are
the light-cones of points P and Q in regions I and II that open toward
\jmin, and null (``Y'') and timelike (``X'') geodesic segments with an
endpoint at $x_i$ and crossing \jmin.
\label{fig-conform}}
\end{figure}

\subsection{Eliminating the beginning}

The above semi-eternally inflating model can be made eternal by
setting the ``state'' of the universe to be exactly that state {\em
approached} by semi-eternal inflation: because the statistical
properties depend only upon $t-t_0$, for specified conditions at $t_0$ the
state at fixed $t$ with $t_0 \rightarrow -\infty$ is the same as that
for $t\rightarrow \infty$ with fixed $t_0$.

The state so obtained has the four basic characteristics listed above:
The spatial sections are exactly flat (outside of the bubbles), the
bubble distribution (as characterized by the incoming bubble rate) is
homogeneous and independent of time, as is the distribution of
inflating regions about any inflating region, and the bubbles are all
statistically identical. The inflating region is a fractal of
dimension $D < 3$ on all scales.  This means that, although inflating
regions exist, the global inflating fraction $f_{\rm inf}$ is zero,
just as the fraction of 3D Minkowski spacetime filled by an infinite
2-plane of finite thickness---an object of fractal dimension
two---would vanish. (The zero probability that a randomly chosen point
is in an inflating region accords with the fact~\cite{ssei} that within the PLC of
each point there is an infinite 4-volume in which bubbles can nucleate
toward that point.)  Unlike the region filled by the plane, however,
the inflating region is statistically homogeneous and isotropic, in
that it exactly satisfies the ``Conditional Cosmographic Principle''
of Mandelbrot that the statistical description of the inflating region
about any given inflating region is independent of the inflating
region chosen (see Ref.~\cite{mandel} for a discussion of this and
other aspects of ``cosmological'' fractals).

This model (essentially derived by Vilenkin~\cite{vil92}) would seem
to have exactly the properties expected of an eternally inflating
spacetime, has been straightforwardly constructed using the
steady-state generated by a semi-eternally inflating model, and
extends to infinite negative cosmic time.  Yet the arguments
of~\cite{singth,bgv} (discussed in more detail in Sec.~\ref{sec-aot})
imply that it should be geodesically incomplete.  This issue can be
addressed with reference to the conformal diagram of the model, shown
in Fig.~\ref{fig-conform}.  The background inflating spacetime is
represented by the lightly shaded region.  Each equal-time surface is
intersected by an infinite number of bubbles (indicated by light-cones
opening toward ${{\cal J}^+}$), which are concentrated along ${{\cal
J}^+\,}$ and near $i^0$.

The model thus far constructed (defined in the shaded region
henceforth called ``region I'') is geodesically incomplete: all null
geodesics (such as ``Y'' in Fig.~\ref{fig-conform}), and all timelike
geodesics (such as ``X'') other than the comoving ones, have only a
finite proper time (or affine parameter) between a point in region I
and coordinate time $t\rightarrow -\infty$ (see, e.g.,~\cite{ssei}).
Most geodesics thus ``leave the spacetime'' to their past,
encountering \jmin, the limit surface of the flat equal-time surfaces
as $t\rightarrow-\infty$. (On the conformal diagram this surface looks
like a null cone emanating from a point at the bottom edge.)

	Although the spacetime is geodesically incomplete there is no
curvature blowup or other obvious pathology at \jmin, so the spacetime
is extendible rather than singular.  One may take the position that
this sort of incompleteness is allowed, since the edge is outside of
the future of any point in the region, and any given thing in the
spacetime was made at some particular coordinate time $t >
-\infty$\footnote{This was the view taken by investigators of the
``cyclic model''~\cite{cyclic} which is geodesically incomplete in a
very similar way, and may be the view taken by the adherents of the
classic steady-state model.}.  From this point of view, there is no
clear reason to reject the model as defined in region I.

It seems quite reasonable, however, to ask instead how the manifold
could be extended, and what could be in the extension.  We start by
extending the manifold to include \jmin, which is the boundary of the
open set comprising region I.  We shall see, as follows, that on \jmin
the field must everywhere be in the false vacuum.  Define
$\phi(\lambda)$ as the field value at affine parameter $\lambda$ of a
non-comoving geodesic starting at some arbitrary point $x_i$ in region I,
where $\lambda$ increases away from \jmin.  We know that for some
affine parameter $\lambda_{\cal J}$, the geodesic encounters \jmin,
and also that if our point is within a bubble, there is also a finite
value $\lambda_f > \lambda_{\cal J}$ at which the geodesic leaves the
bubble and enters the false vacuum $\phi_f$. Then for $\lambda <
\lambda_{f}$, we have
\begin{equation}
\lim_{\lambda\rightarrow\lambda_{\cal J}} \phi(\lambda)=\phi_f.
\end{equation}
If we then require that the field be continuous along any geodesic, we
then find that the field must be in the false vacuum $\phi_f$
everywhere on the surface \jmin.  That is, at the semi-classical level
of description, \jmin must be an infinite null surface of pure false
vacuum, through which no bubbles pass.

	Let us now examine the global classical structure of the
background spacetime by momentarily neglecting semi-classical
processes such as bubble nucleations.  Then the manifold comprised of
region I and \jmin is locally dS (constant Ricci scalar $R$ and
vanishing Weyl curvature tensor) everywhere, with $\phi$ in the false
vacuum ($\phi=\phi_f$).  This manifold can still be extended past
\jmin, and the obvious extension is to complete dS spacetime.
This is certainly {\em a} solution compatible with our state at \jmin,
and (as we will argue in Sec.~\ref{sec-bcs} and the Appendix) it seems
likely to be unique. Thus we will take the maximal extension of the
background spacetime to be full dS spacetime.  That is, the non-shaded
region of Fig.~\ref{fig-conform}, henceforth called ``region II'' must
simply be the rest of dS spacetime.  Consider now a classical field in
the background spacetime obeying a homogeneous hyperbolic equation.
Given any point $P$ in region I, almost all inextendible non-spacelike
curves through $P$ intersect \jmin. Therefore specifying the field
values on \jmin effectively poses a ``characteristic initial value
problem''~\cite{courhil,penrose} with a unique solution everywhere in
region I (this is the analog of the Cauchy problem, but with boundary
conditions on a null surface; see Sec.~\ref{sec-bcs} and the Appendix
for more details).  Exactly the same argument can be made, however,
for any point in region II.  Thus specifying classical fields
everywhere on \jmin determines their values everywhere in dS
spacetime. This means that the conditions found to obtain on \jmin (by
specifying the state in region I and requiring fields to be
continuous) {\em also} determine the state in region II and we can
extend our model to region II in an essentially unique way.

We may now examine the extension of the model to region II at the
semi-classical level by including the bubble nucleations.  The form of
$U(\phi)$ indicates that bubbles must nucleate\footnote{Bubble
nucleation has been perhaps most rigorously analyzed in
Ref.~\cite{rubsib}, and the boundary conditions for bubble nucleation
in our model correspond exactly to those analyzed in their study.} at
a fixed rate per unit physical 4-volume.  In region I, this led to an
asymptotically steady-state bubble distribution which, when made
exact, implied that there are {\em no} bubbles passing through \jmin.
Thus in region II, though bubbles must nucleate at the required rate,
none must pass through \jmin.  

\begin{figure}
\includegraphics[width=8.5cm]{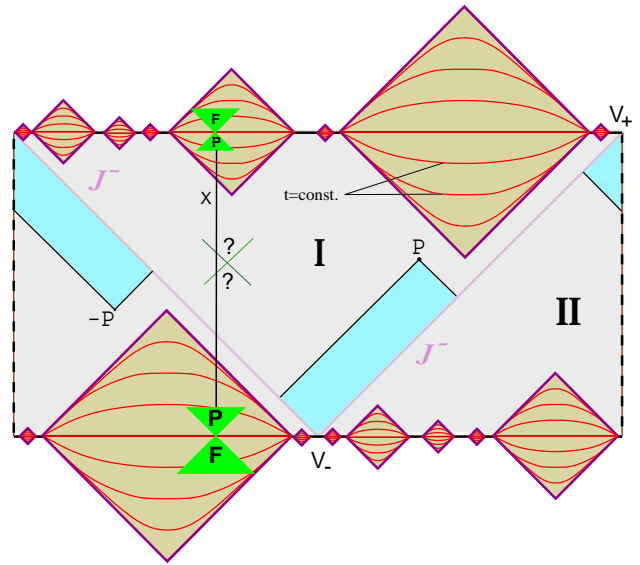}
\caption{Conformal diagram for eternal double-well inflation.  Bubbles
are open FRW regions; equal time slices shown as curved horizontal
lines. For clarity we have not included bubble intersections.  Also
shown are past light-cones, cut off at \jmin, of both a point $P$ and
its antipode $-P$ (note that $P$ and $-P$ are also reflected across
the suppressed two-spheres in the 4D case).
\label{fig-fullmod}}
\end{figure}

The only way this may occur is, in exact symmetry with region I, to
have a steady-state bubble distribution on the flat slices of region
II, with the bubbles opening {\em away} from \jmin.  This is illustrated in
Fig.~\ref{fig-fullmod}.  Region II is, then, a sort of mirror
reflection of region I through \jmin.  Thus the answer to the question
of how the model can be extended, and what lies beyond \jmin, is that
an essentially identical copy of region I lies in region II, connected
by the infinite null surface \jmin.

	This completes the basic specification of our model.  The
cosmology obeys the CP and PCP in that it admits a coordinatization
such that all spatial slices are statistically homogeneous, and
statistically the same as all others.  There is no preferred time in
this slicing, nor is there a cosmological singularity: the model
is geodesically complete.  Given the appropriate inflaton potential,
any one of the bubbles could describe our observable surroundings.

\section{The arrow of time}
\label{sec-aot}

	If we consider all bubbles to expand with time, then
Fig.~\ref{fig-fullmod} suggests that while in region I the future is
toward the top of the diagram (``up''), in region II future lies
toward the bottom (``down'').  This leads us to the issue of the
cosmological arrow of time (AOT): why does the time-asymmetric 2nd law
of thermodynamics hold universally, given that fundamental physics is
thought to admit a symmetry (CPT) that includes time-reversal? There
is some consensus that if this question has an answer, it must
ultimately be cosmological, with the time asymmetry resulting from
some qualitative difference between cosmological ``initial'' and
``final'' conditions~\cite{aotbook} that precludes a time-reversal
(T-) symmetry of the physical state in any sub-region of the universe,
and hence induces an AOT.
	
	Although our our model has no ``initial'' conditions, it does
have boundary conditions on \jmin (discussed in detail in
Sec~\ref{sec-bcs} below) and we can discuss the AOT in light of
them. To do so we must divide the universe into two types of
sub-regions: those entirely outside of bubbles, and those partially or
wholely within them.  Outside of the bubbles (or alternatively near
enough to \jmin) there {\em is no local AOT}: the description of such
a region admits a time-reversal symmetry.  Were we to hypothesize, for
example, an imaginary observer outside of a bubble with its own AOT
(pointing away from the time of its creation), that observer would see
only T-invariant dS spacetime.  The observer could not know whether it
moved ``up'' or ``down'' on the conformal diagram, nor if it was in
region I or II, nor if it crossed \jmin.  What the observer is
guaranteed, however, is that it will eventually be encountered by, and
find itself within, a bubble.

	The bubble interiors are not time-symmetric: within a bubble,
there is a unique time direction in which the mean energy density
decreases.  This direction is away from the bubble wall/slow-roll
inflation epoch, at which the FRW-region is known to be nearly
homogeneous.  If one bubble is to represent our observable
surroundings, this direction must correspond to the time direction in
which the entropy of an isolated system increases.  It has been often
argued, particularly by Penrose~\cite{penroseic}, that this connection
arises because when gravity is included homogeneity corresponds to an
extremely low-entropy state.  We shall assume this correspondence here
(and that the bubble does not begin in some very special state for
which the density fluctuations decrease).  Under this assumption the
physical AOT within any bubble must point {\em away} from the bubble
walls; globally this means that the AOT (where defined) points away
from \jmin.

	As illustrated in Fig.~\ref{fig-fullmod}, one can therefore
indeed draw timelike geodesics (such as ``X'') along which the
physical AOT reverses, but the reversal always occurs to the past of
any physical observer (all of which are within bubbles), and within a
region (the locally dS spacetime) in which there is no well-defined
physical AOT.

	We have argued that within bubbles the physical laws, but not
the physical state, admit a symmetry (CPT) including time-reversal,
while outside of bubbles the laws {\em and} the state admit time
symmetry; but what about the bubble nucleations themselves?  Is there
not some AOT telling them ``which way'' to nucleate, depending upon
which side of \jmin they are on?

\begin{figure}
\includegraphics[width=8.5cm]{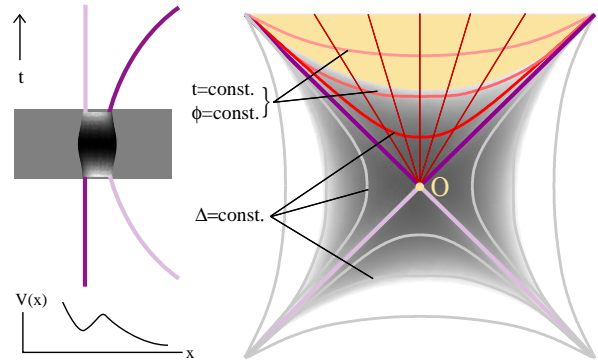}
%\vskip0.15in
\caption{The decay of: (left) an unstable particle, (right) an
unstable vacuum state.  In the particle case, the trajectory is
classically describable to good precision at early and late times, but
not near the decay (shaded region).  Likewise, the inflaton $\phi$ is
classically describable at large invariant distance $\Delta$ from the
nucleation event at $O$, but not near it (shaded region). This quantum
region connects the classically describable field configuration of the
bubble interior to that of locally dS spacetime.
\label{fig-bubbnuc}}
\end{figure}

To clarify this point, consider the process of semi-classical bubble
nucleation by analogy to the decay of an unstable particle, described
by a potential $V(x)$ as in Fig.~\ref{fig-bubbnuc}.  Classically, one
allowed trajectory is for the particle to sit at constant $x$ in the
false minimum $x_f$ (vertical line on diagram); a second, equal-energy
trajectory is for the particle to ``bounce'' off of the potential
(curved line in diagram).  A purely quantum description of the system
(in the Schr\"odinger representation) would start, for example, with a
Gaussian wave function centered at $x_f$ at some time $t=0$, and hence
with the position expectation value $\langle \hat x \rangle \simeq
x_f.$ With time, the wave function spreads, and $\langle \hat x
\rangle$ increases, eventually approaching the classical trajectory
for $\langle \hat x \rangle \gg x_f$ (we could say roughly that the
particle has ``decayed'' when $\langle \hat x \rangle$ changes
significantly from $x_f$.)  Note also that the same spreading would
occur at $t < 0$ so that this purely quantum description would be
T-symmetric.  Now, a {\em semi-classical} description of the system
would describe the system classically at both early and late times,
but with a quantum-mechanical transition connecting the classical
trajectories at some given time. This transition time is random, and
in an ensemble of such systems (as is required for a correct
probabilistic description) would follow a probability distribution
given by a WKB-type calculation of the decay rate.  Near the
transition time the system cannot be described classically; we must
``shade out'' the region where only a quantum description is accurate,
as in Fig.~\ref{fig-bubbnuc}.

Bubble nucleation can be described semi-classically in a similar
way. Here, we must attach allowed classical solutions of the field
equations along some boundary that represents the nucleation event.
To do this in a covariant way, this boundary must be a surface of zero
proper distance, i.e.\ a null cone, as shown in Fig~\ref{fig-bubbnuc}.
A bubble nucleation ``event'' is thus comprised of a region (shown as
the shaded upper quadrant) where a classical bubble interior solution
applies, attached to locally dS regions by a ``shaded out'' region
(within some proper separation squared of the nucleation point) where
only a quantum description is valid.  To produce a semi-classical
description of a spacetime in which nucleation events occur, one must
then populate it with these configurations in such a way that the
nucleation sites are randomly situated and occur at the correct rate
per unit four-volume, while the classical description, which applies
far from the nucleation site, is in accord with the (classical)
boundary conditions.  When the classical boundary conditions are ours,
given on \jmin, this yields the bubble distribution indicated in
Fig.~\ref{fig-fullmod}.  The (semi-classical) boundary conditions do,
then, control the time direction of bubble nucleation, not by
introducing some locally-detectable AOT, but by controlling the
allowed global configuration of bubble nucleation events.

	There is one final ``region'' in which we can check the AOT:
the entire universe.  Interestingly, we here find that while each
bubble nucleation event is non-time-symmetric, by virtue of the
symmetry of the cosmological boundary conditions, the statistical
description of the universe does admit a sort of T-symmetry. In
Sec.~\ref{sec-elliptical} we will discuss the possibility of making
this symmetry exact via an identification on the manifold.  This
raises the intriguing possibility of having a well-defined (and
consistent among communicating observers) AOT for all observers even
while the physical laws {\em and} the global physical description of
the universe both admit a time-reversal symmetry.

\subsection{The singularity theorems}

	Having examined the AOT, we may turn to the singularity
theorems, for the AOT proves crucial in their analysis.  Both the
older theorems~\cite{singth} and the newer theorem~\cite{bgv} assert
that if certain conditions are satisfied everywhere within a spacetime,
then not all past non-spacelike geodesics have infinite proper time or
affine parameter.  This indicates that the spacetime is either
extendible or contains singularities.

The older theorem poses four such conditions, of which our model
satisfies three, as does region I or II by itself.  The fourth
condition is that there exists a point $P$ such that there is a finite
difference in volume between the interior of the PLC of $P$ and that
of any point $P'$ in the past of $P$.  This is motivated by
double-well inflation, and claimed as necessary for semi-eternal
inflation (the argument being that if this condition does not hold,
then any inflating point will find itself in a bubble at the next
instant, with probability one).  We argue that this condition is not
quite necessary: what is required is that there be a finite volume in
the region between the PLCs of $P$ and $P'$ {\em in which bubbles can
nucleate toward $P$}.  Thus if, as in the proof of the theorem,
the``past of $P$'' is taken to be the full volume interior to the
light-cone pointing in the time direction away from which a bubble
nucleated at $P$ would expand, then the fourth condition applies to
region I alone (and correctly implies that it is extendible). But it
would not apply to the full spacetime (which is neither extendible nor
singular), because the relevant part of the light-cone extends only to
\jmin.

The argument of the newer singularity theorem~\cite{bgv} consists of
the definition of a local ``Hubble parameter'' ${\cal H}$ meant to
represent the rate of divergence of neighboring comoving test
particles, along with an argument that any region in which a suitable
average ${\cal H}_{\rm av}$ of ${\cal H}$ is greater than zero along
all geodesics must be geodesically incomplete.

We understand Borde at al.'s argument as follows.  One imagines some
timelike or null test geodesic with affine parameter $\lambda$ in the
spacetime in question, then attempts to construct a timelike vector
field $u^\mu (\lambda)$ along the test geodesic into its past such
that ${\cal H}_{\rm av}$ (defined via $u^\mu$) exceeds zero.  It is
shown that this can only be achieved along some finite affine length
of the test geodesic, since the imposed condition rapidly forces
$u^\mu$ towards nullness.  Borde et al.~\cite{bgv} then take their
result to mean that an eternally inflating spacetime is
past-geodesically incomplete.

We take the hypothetical satisfaction of their averaged Hubble parameter
condition for all test geodesics as the implicit definition of what
Borde et al.\ mean by an eternally inflating spacetime.  The logic is
that $u^\mu$ might be independent of test geodesic and simply be the
velocity field of some set of comoving worldlines in the inflating
spacetime.  So what the theorem actually implies is that is it
impossible to entirely cover a spacetime with such a set of worldlines
in a way that allows all test geodesics cutting these worldlines to
obey the Hubble parameter condition.

For illustration, let us consider dS space.  We note that as pure dS
is a maximally symmetric spacetime, it does not make sense to regard
some parts as inflating and others not---one must break the symmetry
by adding some other ingredients, and then frame a discussion in terms
of them.  Nevertheless, let us investigate if the geodesics defined by
having fixed comoving coordinates themselves constitute a $u^\mu$
field satisfying ${\cal H}_{\rm av} > 0$.  It turns out that ${\cal
H}$ reduces to the usual $\dot{a}/a$ in this situation.  In the closed
slicing with metric
\begin{equation}
ds^2=-d\tilde t^2 + H^{-2}\cosh^2 H \tilde t \(d\chi^2+\sin^2\chi
d\Omega^2_2\), 
\labeq{closedds}
\end{equation}
the coordinates cover all of dS, so $u^\mu$ is globally defined.
However, ${\cal H} < 0$ for $\tilde t \leq 0$ so ${\cal
H}_{\rm av}$ goes negative there.  Now consider using a single flat or
open coordinate patch to define $u^\mu$. ${\cal
H}_{\rm av} > 0$ here (at least for the appropriate choice of time
orientation).  However, because neither of these coordinate patches
covers the spacetime, neither does $u^\mu$.  Furthermore, one may
choose an infinite number of flat coordinatizations of dS, each with a
different null boundary where the $u^\mu$ construction fails.  This
makes it clear that the boundary of a given $u^\mu$ field cannot be
unambiguously used to define an edge to an inflating region.

We do not believe that the global existence of a suitable $u^\mu$ is
necessarily the best definition of what is meant by an eternally inflating
spacetime.\footnote{Indeed the term ``eternal inflation'' has been
used with a variety of meanings. For example, the recent paper~\cite{em}
used it to describe models that are eternally inflating to the future,
but simply geodesically complete to the past ($\dot a$ and/or $\ddot
a$ may go negative there).  Such histories, with a globally-defined
arrow of time, seem physically unrealistic with a typical mechanism
for exiting inflation, since this would render them unstable to the
formation of thermalized regions in the putatively eternal early
phase\cite{etin}.} In particular, equating ${\cal H}$ with physical
expansion entails a tacit assumption that the physical AOT is
everywhere in accord with that defined by $u^\mu$.  The model we have
proposed could be covered by a congruence of geodesics (those comoving
in two flat coordinate patches covering dS) that would yield ${\cal H}
<0 $ in some regions.  However, we have argued that these regions may
still be regarded as expanding with respect to the physical AOT
defined by the cosmological boundary conditions.

In summary, both singularity theorems postulate conditions for a
region to be ``inflating'', and find that such a region cannot be
geodesically complete.  However, interpreting these theorems as
forbidding eternal inflation seem to us to require an unwarranted
assumption about the global AOT independent of the cosmological
boundary conditions.

\section{Boundary conditions and the null boundary proposal}
\label{sec-bcs}

We have seen how extending semi-eternal inflation to eternal inflation
implies particular behavior on the infinite null surface \jmin.  Here
we discuss the converse, describing how the eternal double-well
inflation model we have described can be specified by a particularly
simple set of cosmological boundary conditions that are imposed on
\jmin.

\subsection{On cosmological ``initial'' conditions}

	The correct specification even of a complete set of physical
laws does not by itself allow prediction of any physical system's
behavior; these laws must be supplemented by boundary conditions that
suffice to fully characterize the system being modeled.  The Big-Bang
(``BB'') model essentially consists of a set of such boundary
conditions for our observable cosmic surroundings: at some early time,
our region was a hot, dense, nearly homogeneous and isotropic mixture
of particles and fields in thermal and chemical equilibrium, in a
nearly-flat expanding background geometry with scale-invariant
gravitational potential perturbations of amplitude $10^{-5}$ on the
scale of the cosmological horizon.

	While these initial conditions yield predictions in excellent
accord with astronomical observations, they are deemed by many
to be too special: they seem to comprise an extremely small
portion of some ``ensemble of all possible initial conditions'', using
some (generally unspecified) measure (see e.g.~\cite{inflnotperf}).
The theory of inflation is widely accepted as a way to broaden the
range of allow initial conditions by funneling a relatively wide class
of physical conditions into satisfactory BB initial
conditions~\cite{linde92}.

This is perhaps reasonable as a stop-gap measure as long as the true
(pre-inflation) initial conditions are unknown.  But it does not {\em
solve} the initial condition problem, for not all initial conditions
will give rise to inflation~\cite{vasptrodd}, nor is it clear that all
of those that do give inflation will yield a viable big-bang
model~\cite{inflnotperf}. Thus one must still {\em assume} that
suitable conditions emerge from the initial singularity; whether this
assumption corresponds to a ``special'' or ``generic'' condition seems
ill-defined without a description of the singularity, which would
presumably require quantum gravity.  But unless quantum gravity
completely alters the way physical theories are applied, it is
unlikely to yield a {\em unique} initial condition; it will still be
necessary to supply a quantum state (or something similar).  Even if
it is possible to define the ensemble of all possible states, the
question of how (and why) a state was ``chosen'' for our universe can
only be a metaphysical one, and it seems that, just as in choosing
physical laws, we can at best posit a state on some (possibly
symmetrical or aesthetic or philosophical) grounds, derive its
consequences, and compare them to the observed universe.

	The same holds for a non-singular cosmological model such as
that proposed in this paper, the basic structure of which is
well-described by semi-classical physics, and where there can be no
hiding of the cosmological boundary conditions in an initial
singularity. We instead specify a particular set, and analyze
them in terms of their simplicity, symmetry, and ability to correctly
generate a big-bang like region that can describe our observable
surroundings.

\subsection{(Semi-)classical boundary conditions for eternal inflation}

	It is conventional to pose boundary conditions for a set of
classical fields by specifying the fields (and generally their time
derivatives) on a spacelike surface. This accords with the intuitive
idea of specifying the state at an initial time. An alternative
procedure, more relativistic in spirit because it does not assume a
particular time coordinate, is to specify boundary conditions on a
null surface such as the light-cone of a point. (See the Appendix and
Refs.~\cite{courhil,penrose,pr} for treatments of the null initial
value problem.)

	Our boundary surface \jmin is a null surface that, when drawn
on the conformal diagram, looks like the light-cone of a point $V_-$
at the bottom (i.e. ``at'' the conventional past timelike infinity)
{\em and} the light-cone of a point $V_+$ at the top; see
Fig.\ref{fig-fullmod}).  As such, one might expect that one may
determine fields (including the spacetime metric) throughout the space
by specifying them on \jmin.  The conformal mapping, however, hides
the fact that \jmin is not a fully closed light-cone: there is always
a nonzero physical volume (of order $H^{-3}$ or greater) on any
spacelike surface bounded by \jmin. We must therefore ask if
specifying fields on \jmin suffices to fix them everywhere, or if
there is information that may come ``though the hole'' at $V_-$ and
$V_+$.  In the Appendix, we argue (using the Green function for a
massive scalar field in dS) that the fields at $V_\pm$ are irrelevant
(as long as they are reasonably well-behaved), as their effect is
infinitely diluted.

We can thus pose boundary conditions for our cosmology at the
classical level as:
\begin{enumerate}
\item{There exists an infinite connected null hypersurface \jmin of
topology $R \times S^2$, on which the 4-dimensional Weyl tensor
vanishes and the 4D curvature scalar is constant.}
\item{The inflaton field, with a ``double well''
potential, is everywhere in the false vacuum on this surface.}
\item{On this surface, all classical fields are zero or are in minima
of their potentials. This precludes any radiation propagating through
\jmin.}
\end{enumerate}

This cosmology is classically very dull, as it is just de Sitter space
everywhere with no dynamics.  However, semi-classical bubble
nucleations can, without affecting the fields on \jmin, create
interesting dynamics by forming bubbles that open everywhere away from
\jmin, and give rise to eternal inflation as described in
Sec.~\ref{sec-prop} and shown in Fig.~\ref{fig-fullmod}.

\subsection{Quantum mechanical boundary conditions}

	Although a fully quantum description of semi-classical
phenomena---such as the nucleation of a bubble---is generally
prohibitively complex, we may hope that, because our set of boundary
conditions is  so simple at a semi-classical level, it might be amenable
to a simple quantum formulation, and it would be pleasing to specify
our boundary conditions in an explicitly quantum-mechanical way.
There are two ways one might think of doing this.

First, one might consider doing usual quantum field theory on de
Sitter space, but with a particular choice of vacuum corresponding to
our ``false vacuum'' classical conditions on \jmin.  This might be
done by putting the field in a Gaussian wavepacket around the
false vacuum at (flat slicing) time $t$, as in~\cite{rubsib}, then
taking a $t\rar -\infty$ limit. One might then assume that this is
sufficient to define the quantum state over all of de Sitter space.
Alternatively, one could set such an initial state at the $\tilde t=0$ slice
in global (closed) coordinates~\eqn{closedds}, i.e. on the slice
through the ``throat'' of the dS hyperboloid.  This surface could then
be boosted (infinitely) to become \jmin.  In either case, the
procedure would be analogous to putting an unstable particle into a
Gaussian in the false vacuum at $t=0$; away from $t=0$ in both time
directions, the solution of the time-dependent Schr\"odinger equation
would evolve away from the unstable ``initial'' state.

With our initial data surface being null, one might attempt to define
the quantum state directly on it.  This would be conceptually more
pleasing, not requiring the limit through a sequence of spacelike
surfaces as above.  Further, a null surface formulation of initial
data is rather elegant~\cite{penrose,pr}.  However, unlike for a
spacelike initial value surface, some points on \jmin are in causal
contact.  
It is thus less easy to see how to move from a
spacelike to a null boundary value surface in QFT, because fields do
not commute at lightlike separated points.  A similar effect occurs in
so-called ``light-cone quantization'' approaches to QFT in Minkowski
space~\cite{heinzl}, where one uses infinite null sheets (or
lightfronts) instead of equal time slices, and fields again do not in
general commute.  The vacuum state for interacting fields is rather
easier to define in the lightfront approach to QFT than in the usual
spacelike approach to QFT, and one might expect this to hold in a
proper null {\em cone} approach to QFT.  Unfortunately, the authors
know of no such formulation of QFT even in Minkowski space.

Thus while we suspect that our cosmological boundary conditions may
correspond to a the specification of a rather simple quantum state, we
leave this difficult problem for future work and here concentrate on a
semi-classical description (though we return to QFT in
Sec.~\ref{sec-elliptqft}).

\subsection{Discussion of the null boundary proposal}

	The boundary conditions we have proposed are extremely
simple, in line with the view espoused above that one cannot avoid
making a specific choice for cosmological boundary conditions, and
that it is then reasonable to make a choice that is as simple and as
highly symmetric as possible, rather than hope to choose a ``generic''
boundary condition.

\subsubsection{Relation to other proposals}

Our proposal is novel in that it requires no treatment of an initial
singularity, and in that boundary conditions are placed on a null
boundary that does not correspond to any particular cosmic ``time''.
It does, however, have features in common with other proposals for
specifying the state of the universe.

Penrose's Weyl Curvature Hypothesis (WCH) requires that initial
singularities are constrained to have vanishing Weyl tensor,
corresponding to a low-entropy state.  This, he argues, gives rise to
an arrow of time flowing away from the initial state.  Our proposal is
very similar, except of course that we impose the vanishing of the
Weyl tensor on a lightlike slice running across the universe rather
than at an initial singularity.

The Hartle-Hawking No Boundary Proposal (NBP) is a prescription for
the wavefunction of the universe.  It provides $\Psi (h_{ij},\phi)$,
the amplitude for a given three-metric and scalar field configuration
on a spatial 3D surface.  The construction is designed to suppress
irregular configurations relative to more regular ones, and to thus
favor simple and symmetric states.  The notion of temporal evolution
does not appear explicitly.  However, one can associate histories with
saddle-point approximations to the wavefunction, and it turns out that
such universes are often smooth when they are young and small.  Thus
the NBP may be taken to imply that, at the semi-classical level, there
should exist a spatial surface in a cosmological spacetime on which
the boundary conditions are particularly simple.\footnote{In the
context of models with open-inflationary potentials, the NBP seems to
favor histories in which the scalar field is everywhere in its true
vacuum~\cite{hommode}.  The NBP may still be relevant for inflation
with the use of an anthropic constraint, or in a ``top-down'' approach
to calculating quantum probabilities~\cite{hawkhert}.  Starting off in
the true vacuum is however not a problem for recycling models of
inflation, as discussed below.}. Our proposal rather uses a null
surface (but see Sec.~\ref{sec-diffbound} below).  It would be very
interesting to develop a quantum prescription in a similar vein to the
NBP which naturally leads to simplicity on certain null surfaces.
For further details on both the WCH and NBP
see~\cite{natspace}.

The tunneling approach to quantum cosmology~\cite{tunnel} argues
that the semi-classical universe emerges via a quantum tunneling
event.  In the context of models with open-inflationary potentials the
proposal suggests that the universe, when first semi-classically
describable, is most likely to be small and regular, with the field
away from the true vacuum; future-eternal inflation can then ensue.
The tunneling is supposed to have occurred out of a quantum
gravitational chaos so severe as to preclude any space-time
description.  While our proposal also leads to inflation from
semi-classical boundary conditions, it explicitly avoids any such
extreme quantum gravitational regime.

Over the years Sakharov has discussed various cosmological models
involving time-reversal (and CPT)
invariance~\cite{sak1,sak2,sak3,sak4}.  In these models, the universe
is generally symmetric across a singular FRW bounce at $t=0$, at which
universe assumes an especially simple state; away from this bounce
entropy increases in both directions of time~\cite{sak2}.
In~\cite{sak1} he hypothesizes that phenomena at $t<0$ are the CPT
reflections of the phenomena at $t>0$.  In~\cite{sak2,sak3} he
considers the possibility of an infinite chain of further bounces or
cycles away from $t=0$ in both time directions, and also the
possibility that the minimum-entropy surface might be one of maximum
expansion rather than a singular one~\cite{sak2,sak3}.  In a paper
primarily about signature change~\cite{sak4} he alludes to possibility
of non-singular time-reversal around a surface of minimal radius in a
false vacuum state. Our proposal clearly has parallels with these
ideas.  We, however, concentrate on an infinite non-singular null
surface, rather than a singular spacelike one.  Moreover, with the
concepts of semi-classical bubble nucleation and open inflation, we
are able to provide a relatively complete physical picture.

\subsubsection{The horizon problem}
\label{sec-horiz}

	Inflation was originally conceived as a remedy for troubling
issues concerning cosmological initial conditions~\cite{guth80}, thus
it is useful to compare how such shortcomings of the HBB
model~\cite{linde92} are dealt with in our model.  We focus on what is
(aside from the singularity problem) perhaps the most vexing of these
HBB difficulties: the horizon problem.

The horizon problem is generally framed as follows: choose two
spatially antipodal points on the last-scattering surface.  They are
similar in temperature, yet their PLCs never intersect in a HBB
cosmology, so there can be no causal connection between them.
Inflation is generally thought to solve this problem, because with
sufficient inflation, the PLCs will intersect.  But this does {\em
not} suffice for the points to have the same temperature, because the
temperature at each point depends on data across its full PLC, and
there is a portion of each PLC that does not intersect the other.
Taking this into account, for the two points to have similar properties
it must be assumed that there is sufficient inflation, and
additionally that at inflation's beginning at time $t_{\rm inf}$, the
region to the past of the two points is homogeneous on length scales
of order $H^{-1}_{\rm inf}$.  But suppose there is a earlier epoch
between some time $t$ and $t_{\rm inf}$.  Then at $t$ the patch must
be homogeneous over a physical length scale of at least
\begin{equation}
r(t) \sim {a(t)H(t)\over a(t_{\rm inf})H(t_{\rm inf})}H^{-1}(t),
\end{equation}
where $H(t)\equiv \dot a/a$.  If $a\propto t^\alpha$ ($\alpha=1/2$ for
radiation-dominated expansion) prior to $t_{\rm inf}$, then
$r/(H^{-1}) \simeq [a/a(t_{\rm inf})]^{(\alpha-1)}$.  Thus it is
necessary to postulate that our region was, at some initial time
$t_0$, homogeneous over $[a(t_{\rm inf})/a(t_0)]^{(3-3\alpha)}$ Hubble
volumes.  The horizon problem therefore persists if $t_{\rm inf} >
t_0$ and $\alpha < 1$. There are two escapes available within
inflation.  The first is to set $t_{\rm inf}=t_0=t_{\rm pl}$, so
that the expansion is inflationary all the way back to the Planck
time, before which one cannot speak of the expansion at all. The
horizon problem is then greatly ameliorated, as it must only be
assumed that some regions ``emerging'' from the Planck epoch are
homogeneous over a relatively small (but $> 1$; see~\cite{vasptrodd})
number of Hubble volumes when they are first classically describable;
but whether this constitutes a true solution will be seen only if and
when the Planck epoch itself is understood.  A second potential escape
is to set $\alpha\ge1$, or (more generally) for $\dot a$ to be
non-decreasing, i.e.\ to have past-eternal inflation.

How, then, does the horizon problem look in the context of our eternal
model with no initial time?  Let us pose the problem in a slightly
more general manner: when specifying the cosmological boundary
conditions, must one do so over a region that is very large compared
to some relevant physical volume such as $H^{-3}$?  Posed this way, it
might seem that because \jmin is an infinite surface, a strong horizon
problem exists.\footnote{This is apparently the case, for example, in
the cyclic model~\cite{cyclic}, where if the cyclic behavior is to
continue indefinitely into the future and past, it seems necessary to
place cosmological boundary confitions on an infinite spacelike
surface.  This specification accords the same properties to points
that are arbitrarily distant, and causally disconnected.} This,
however, is not necessarily the case.  Imagine \jmin as the limit of a
sequence of spacelike slices obtained by boosting the spacelike
surface given by $\tilde t=0$ in global coordinates.  Because each
such surface has volume $2\pi^2 H^{-3}$, we might also attribute to
\jmin the same finite invariant volume.\footnote{One can also consider
other ways of taking a limit to \jmin, but this one seems most in
accord with the symmetries of dS.  Another approach to the problem is
to rather consider some distance measure between any two points on
\jmin.  A dS-invariant quantity, in units where $H=1$, between points
$P_1$ and $P_2$ in the embedding space (see Eq.\eqn{dshyp} below) is
given by
$D(P_1,P_2)=-v_1v_2+w_1w_2+x_1x_2+y_1y_2+z_1z_2$~\cite{dsreview}. In
general $-\infty < D < \infty$, whereas for two points on \jmin, $|D|
\le 1$; they may thus be considered ``close''. The relation between
$D$ and geodesic distance is, unfortunately, not always defined. The
points are timelike or null separated, respectively, for $D>1$ or
$D=1$. For $|D|<1$ points are connectable by a spacelike geodesic of
length $\cos^{-1}(D) < \pi$; but for $D \le -1$ there is {\em no}
geodesic connecting $P_1$ and $P_2$, although one does connect $P_1$
to the antipode $\bar {P_2}$ of $P_2$ because $D(P_1,\bar
{P_2})=-D(P_1,P_2)$. Under the identification of antipodal points (see
below), any $P_1$ and $P_2$ are null-separated or spacelike-separated
by a geodesic distance $<\pi/2H$.} Thus our construction would seem to
ameliorate the horizon problem (posed in terms of Hubble volumes
within the boundary condition surface) to approximately the same
degree as does inflation beginning in the Planck epoch, and much
better than does inflation with an early quasi-FRW phase.

\section{The elliptic view}
\label{sec-elliptical}

	The model we have proposed consists of two indistinguishable
regions, each comprising an eternally inflating universe with an AOT
(where defined) pointing away from an infinite null surface which
connects the two regions.  The statistical identity of these
universes, along with lack of a global physical time orientation,
suggests some form of an old idea concerning dS, called ``the elliptic
interpretation''\footnote{The etymology of this term is obscure to the
authors. The elliptic view applied to {\em spatial} sections of dS was
discussed first by de Sitter, who preferred it to the (now)
conventional ``spherical'' view; and cited a letter from Einstein
voicing the same preference~\cite{ds}.  Antipodally identifying in
space {\em and} time was discussed in fond detail by
Schr\"odinger~\cite{schro}.} that would identify the two universes.

	The idea consists of deeming an event to be represented not by
a single point of a spacetime manifold, but by a {\em pair} of
antipodal points (defined below).  This corresponds to a topological
identification that, applied to our model, identifies regions I and
II, and maps \jmin onto itself (the $R\times S^2$ manifold \jmin
becomes $(R\times S^2)/Z_2$). This identification has been subject of
some previous~\cite{schro,schmidt,gibbons,elliptqft} and
recent~\cite{verlinde} investigations.

	Pure dS can be represented as a 4D hyperboloid
\begin{equation}
-v^2+w^2+x^2+y^2+z^2=H^{-2}
\labeq{dshyp}
\end{equation}
embedded in 4+1 D Minkowski space with metric
\begin{equation}
ds^2=-dv^2+dw^2+dx^2+dy^2+dz^2.
\end{equation}
The elliptic interpretation consists of identifying each point $P$
with coordinates $(v,w,x,y,z)$ with its antipode $-P$ at
$(-v,-w,-x,-y,-z)$. In the conformal diagram, this means that points
such as $P$ and $-P$ of Fig.~\ref{fig-fullmod} are physically
identified; the antipodal map ${\cal A}$ looks like a vertical
reflection and a horizontal shift through one half of the diagram's
horizontal extent (Note that although this map makes an orbifold of
the embedding space, there are no fixed points in the dS hyperboloid,
leaving the identified space a manifold).

At the classical level, the identification can be enforced by
demanding that all fields in dS are symmetric or anti-symmetric under
${\cal A}$, and that all sources have an accompanying antipodal
copy. Parikh et al.~\cite{verlinde} have argued that charge
conjugation should be added to ${\cal A}$, and that the combination
represents CPT.  Indeed, in classical field theories at least, ${\cal
A}$-symmetry automatically entails opposite charges at antipodal
points.  Consider, for example, a complex scalar field, satisfying
$\phi (-P)=\phi (P)$.  Then the global time derivative of the field at
the antipode $-P$ is minus that at $P$, and hence the charge density $
(q /2i) \sqrt{-g} ( \phi^* \partial^0 \phi- \phi \partial^0 \phi^*) $
is opposite. However, in 3+1 D at least, we do not recover the PT part
of their argument.  A particle at $P$ with 3-momentum vector ${\mathbf
p}=(p_x,p_y,p_z)$ has an antipodal copy at $-P$.  By an argument like
that of Parikh et al., parallel transporting the copy's trajectory
back to $P$ takes the momentum to $(-p_x,p_y,p_z)$.  This looks
like parity followed by a rotation of $\pi$ about the $x$-axis.  The
same procedure, however, takes a small displacement ${\mathbf x} = (
x, y, z)$ to $( x, - y, - z)$.  The difference in sign arises because
the 3-momentum suffers an ``automatic'' time reversal under ${\cal
A}$, whereas the 3-displacement does not.  Thus the orbital angular
momentum ${\mathbf l} = {\mathbf x} \times {\mathbf p}$ transforms
from $(l_x, l_y, l_z)$ to $(-l_x, l_y, l_z)$, just as the 3-momentum
does.  But under parity P, $\mathbf{x} \rar -\mathbf{x}$ and $\mathbf
p \rar -\mathbf p$, while under T, ${\mathbf x} \rar {\mathbf x}$ and
${\mathbf p} \rar -{ \mathbf p}$. Therefore under PT (along with
possible rotations), $\mathbf{p}$ and $\mathbf{l}$ transform
oppositely.  Only under T (with rotations) alone do they transform
likewise.  So unfortunately we cannot concur with the beguiling idea
that CPT conjugate events occur at antipodally conjugate points in an
elliptic universe.  Rather, we must settle for a sort of CT
conjugation between processes at antipodal points.

\begin{figure}
\includegraphics[width=8.5cm]{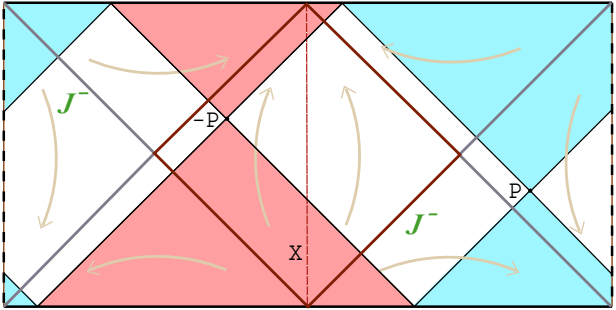}
\caption{Conformal diagram of dS including Killing vectors (arrows), a
point $P$ and its antipode $-P$, along with their light-cones.
The causal diamond of an observer following a
geodesic ``X'' (comoving in the global coordinatization~\eqn{closedds})
is the interior of the dark diamond, the antipodal copy of which is the light diamond.
\label{fig-aotdiag}}
\end{figure}

A second interesting (and favorable) feature of the antipodal
identification, which holds in pure elliptic de Sitter space (edS), is
as follows. An immortal observer in dS can only eventually ``see''
(i.e. be connected to via a non-spacelike geodesic emanating into the
past of the observer) half of the space, the rest being hidden behind
the observer's dS event horizon.  Under the antipodal identification,
however, this ``hidden'' space is exactly the same as the ``visible''
space.  Likewise, the space behind the observer's particle horizon
(the space not reachable by non-spacelike geodesics emanating from an
observer toward its future) is the same as the space within it.  In
this precise sense, edS has no horizons.  The notion that in edS each
observer has ``full information'' about the space has been a prime
motivation for the study of the space~\cite{schro,verlinde}.

Along with these appealing attributes however, and like our
cosmological model, edS has some unconventional temporal features that
may or may not be desirable.  Because ${\cal A}$ includes time
reversal, the spacetime is non-time-orientable: one cannot
continuously divide non-spacelike vectors into two classes which can
be labeled ``future'' and ``past''.  Now, one may take the
view~\cite{schro} that since physics is essentially time-reversible, this
poses no fundamental problem.  Non-time-orientability does, however,
have implications for quantum mechanics (see Sec.~\ref{sec-elliptqft}
and ~\cite{gibbons,elliptqft}).  In addition, while physics may be
time symmetric, our physical world manifestly is not, and this must be
confronted in a cosmological model.

	The identification of points near ``past'' infinity with those
near ``future'' infinity also raises the specter of closed timelike
curves (CTCs) and their accompanying paradoxes.  The identification
does not allow any self-intersecting timelike curves in {\em perfect}
dS because the full light-cones of $P$ and $-P$ never intersect (the
spacetime obeys the strong causality principle of~\cite{he}).  For the
same reason, no observer can see both an event and its antipodal
copy. Note, though, that perturbations of dS that tend to make the
conformal diagram ``taller''~\cite{talltales} would allow timelike
curves from a point to traverse the dS hyperboloid to the point's
antipode.
	
	These temporal features may not be what we are accustomed to,
but perhaps we should not be surprised that any cosmology based on
globally dS spacetime has an ``unconventional'' AOT. It is well known,
for example, that while a patch of dS admits a static
coordinatization, it admits no global timelike Killing vector; as
shown in Fig.~\ref{fig-aotdiag}, the Killing vector is timelike only
in half of the diagram, and moreover points toward the top of the
diagram only in half of the region in which it is timelike, reversing
completely between antipodal points.  (In fact, the Killing vector
field maps onto itself under ${\cal A}$).

	Our model has much the behavior one would want in a cosmology
based on edS.  The physical AOT is defined only inside the bubbles,
which may intersect (and hence compare time orientations) only within
region I or region II.  The two regions are separated by locally dS
spacetime, where physics is fully time symmetric.  Under the antipodal
identification, the regions are equated, and the physical AOT is
consistent everywhere that it can be compared by two physical
observers.  Only an imaginary observer that could travel ``back in
time'' to leave a bubble, pass through \jmin, and encounter another
bubble, could see that its own AOT agreed only with that of one of the
two bubbles.  (In fact, it would seem that in a non-time-orientable manifold
the physical AOT must either be undefined in some regions, or must
suffer a reversal along some surface.  Thus a construction something
like ours may well be necessary in any cosmology based on edS.)

	However, an antipodally identified version of our model does
not quite share all of the desirable properties of edS.  The bubbles,
with a larger curvature radius than the embedding space, allow the
connection of antipodal (and hence identified) points by a timelike
curve.  These self-intersecting timelike curves (SITCs) are not
however CTCs of the usual grandfather paradox sort.  To follow such a
SITC an imaginary observer would, for part of its journey, have to
travel backward in (bubble) time.  Moreover, when the two branches of
the SITC meet, they have opposite time orientation as defined by an
affine parameter along the curve.  One might avoid these SITCs if
the bubbles have a smaller curvature radius than the background space
(as discussed below in Sec.~\ref{sec-general}).  In this case,
however, horizons would return, because there would be regions outside
of an observer's horizon that are not identified with any region
within the horizon.

While the elliptic interpretation is complicated by the presence of
bubbles, the background space of our model is pure dS, and does
benefit from the elliptic interpretation; all points on \jmin, for
example, would be connectable by geodesics and have a maximal spacelike
geodesic separation $\pi/2H$, and the volume of the boundary condition
surface would be halved.

The elliptic interpretation was suggested by a symmetry in the
statistical description of the bubbles, and it is interesting to
ponder the connection between this statistical symmetry and ${\cal
A}$.  Classically, we may have an ensemble of systems each of which is
not ${\cal A}$-symmetric, even while the statistical properties of the
ensemble are.  Without bubble nucleation, our universe (dS) is ${\cal
A}$-symmetric. The boundary conditions which determine the bubble
distribution are ${\cal A}$-symmetric, therefore it seems necessarily
true that the statistics of the bubble distribution are ${\cal
A}$-symmetric. Given quantum mechanics, however, it is not clear how
to relate a single member to the statistical properties (given by the
wavefunction) without bringing in measurement theory, and we will
leave the question aside for future consideration.

\subsection{QFT in edS}
\label{sec-elliptqft}

The non-time-orientability of edS also makes quantum field theory on
the space rather more subtle than in usual dS.  We hope to treat this
subject in some detail in a forthcoming paper (see
also~\cite{gibbons,elliptqft,verlinde}).  Here we sketch a brief
summary.

Consider a massive free scalar field $\phi(x)$ obeying some wave
equation, for which we would like to construct a QFT in edS by
defining a Hilbert space of states (including a vacuum state
$|0\rangle$), and the two-point function $\langle
0|\phi(x)\phi(y)|0\rangle$.  The latter can be decomposed into a
commutator $D(x,y)$ and an anti-commutator (or ``Hadamard'') function
$G^{(1)}(x,y)$. Under the antipodal identification, we would expect
both $D$ and $G^{(1)}$ to be symmetric in some sense under the
exchange of $x$ and/or $y$ with its antipode.  How, then, might we
define such a QFT?  There are a number of ways, none of which seem
entirely satisfactory.

First, one might just pick a particular ``antipodally symmetric''
vacuum state of full dS.  Indeed, taking the $\alpha\rightarrow\infty$
limit of the ``$\alpha$-vacua'' appropriate for dS~\cite{alphavac}
does yield an ${\cal A}$-symmetric $G^{(1)}$.  However, this does not
have the usual short distance behavior of the Minkowski 2-point
function.  In addition, the commutator $D$ is independent of the state
chosen, and has no antipodal symmetry. (One might also hope to find an
${\cal A}$-symmetric non-vacuum state with the correct short distance
behavior, but this would still have the wrong commutator.)
 
A second approach would be to try to build an antipodally symmetric
vacuum in a full dS background, by choosing a (global) time coordinate
and decomposing the fields into global positive-frequency modes that
are antipodally symmetric. The Fock vacuum would then as usual be
the state destroyed by all annihilation operators. The problem with
this approach is that any antipodally symmetric mode turns out to have
vanishing Klein-Gordon norm when integrated over a Cauchy surface for
all of dS~\cite{gibbons}, and the Fock construction breaks down.

A third approach employed in the literature~\cite{gibbons,elliptqft,verlinde}
is to abandon the hope of a Fock vacuum, and to simply enforce the
antipodal symmetry at the level of the Green functions, by writing
antipodally symmetrized versions of the fields, and computing the
resulting two-point functions in terms of the two-point functions of
unidentified dS. One problem with this approach is that it becomes
somewhat unclear why the anticommutator should take this value, as
the construction seems to lack an underlying quantum-mechanical
motivation. Another problem is that while the anticommutator so
obtained is antipodally symmetric, the same procedure yields a
commutator that vanishes identically.

A fourth approach is to define the vacuum in terms of a mode
decomposition over only {\em part} of dS (such as a ``causal
diamond''\cite{dsreview}), where the modes can be consistently
positive or negative frequency (so that a Fock representation exists),
then provide a prescription for defining correlators between any two
points in the space in terms of correlators within this region.  This
approach turns out to be promising; for one or both points in the
causal diamond or its antipodal copy, and for a particular choice of
state (a mixed thermal one at twice the usual de Sitter Hawking
temperature), both $D$ and $G^{(1)}$ can be made to have the correct
symmetry.  The problem arises when both points are outside of the
causal diamond and its copy; in this case the commutator turns out to
vanish for timelike separated points, and not for spacelike separated
points. It is unclear whether this makes sense.

In short, defining a satisfactory QFT in edS is rather difficult; the
difficulties stem primarily from the commutator function, because it
is not symmetric under time-reflection, while the anti-commutator
is. It is then difficult for both functions to be symmetric under
${\cal A}$, which includes time-reflection.  It is
possible~\cite{verlinde} that string theory in edS 
will make more sense than in dS.  It is also conceivable that the
elliptic view could emerge from a correct quantum treatment of dS, and
that the described troubles stem from doing QFT on a fixed background
not included in the dynamics.  This would be an interesting issue to
pursue in string theory or other theories of quantum gravity. For now
we must leave it there, and return to eternal inflation.

\section{Generalizations and extensions}
\label{sec-general}

We have constructed our eternally inflating universe using an ``open
inflation'' double-well potential and demanding that the inflaton
$\phi$ rest in the false vacuum everywhere on the infinite null
surface \jmin.  But these choices are not unique, and the general
principles of our construction can be extended to models employing
different potentials, or different boundary condition surfaces.

\subsection{Different inflaton behaviors}

A simple change in our model can be induced by leaving $V(\phi)$
fixed, but demanding that $\phi$ rest in the true, rather than false,
vacuum on \jmin.  Although the field is now in a (positive) stable
vacuum, bubbles of {\em false} vacuum may still be able to
nucleate~\cite{nucfalse}.  Assuming this indeed occurs, the effect of
each bubble can be enclosed within a light-cone, and the distribution
of these light-cones is essentially the same as for bubbles of true
vacuum, so all of the arguments of Sec~\ref{sec-prop} go through.
Now, within these bubbles of false vacuum new bubbles of true vacuum
can form, one of which could describe our observable universe.  At
late (bubble) times each bubble interior would approach dS, dominated
by the true vacuum (presumably of the magnitude we currently appear to
observe), in which false vacuum bubbles may nucleate (and so on, {\em
ad-infinitum}).  This is a realization of the ``recycling universe''
of ~\cite{recycle}.  Because bubbles are nested infinitely deep, the
structure of this universe appears exceedingly complicated.  But the
{\em global} structure at the outermost level of bubbles is known (and
is as in Fig.~\ref{fig-fullmod}), because we have specified it using
explicit cosmological boundary conditions.  One might argue that this
scenario would have yet simpler boundary conditions than the scenario
in which the inflaton is placed in the false vacuum on \jmin.  Since
the lowest vacuum state must be positive for this scenario to work, it
also has the potential to connect with the presently observed positive
vacuum energy.

A somewhat similar scenario can be realized using ``chaotic
inflation'' potentials such as $V(\phi)=\lambda\phi^4+V_0$.  Here we set
the inflaton to rest at $\phi=0$ on \jmin, and require $V_0 > 0$ so
that the space is locally dS near \jmin. This is, again, a stable
configuration, yet regions of large potential may ``nucleate'' in this
dS background as highly improbable quantum fluctuations in the
field. If such a region nucleates at high enough $\phi$, it would
provide the potential seed for the sort of eternal inflation
envisioned by Ref.~\cite{lindeeternal}, in which quantum fluctuations
in $\phi$ dominate over classical rolling so that inflation become
eternal.  While the structure of the resulting region becomes
extremely complicated, it can again be contained within a light-cone of
some point in the original (dS) embedding space, so that the global
structure of the universe (filled with such light-cones) is still
understandable, and again looks like Fig.~\ref{fig-fullmod}.

\subsection{Different boundaries}
\label{sec-diffbound}

We were led to the null surface \jmin by constructing an eternal model
with a time-translation symmetry.  But the same sort of boundary
conditions we apply on \jmin could also be applied to a {\em
spacelike} initial surface, such as the $\tilde t=0$ surface in the
global coordinatization~\eqn{closedds} of dS.  Such boundary
conditions might be closely tied in to the Hartle-Hawking NBP, as
discussed above. The universe now has a preferred (and initial) time,
and constitutes ``semi-eternal'' inflation in both the $\tilde t < 0$
and $\tilde t > 0$ regions. The same arguments concerning the AOT
apply here: it is undefined near $\tilde t=0$ (outside all bubbles),
and defined in bubbles, pointing away from $\tilde t=0$.  If the
initial surface maps onto itself under the antipodal map, we can apply
the antipodal identification to the universe.  This model bears a
stronger resemblance to the earlier ideas of
Sakharov~\cite{sak1,sak2,sak3,sak4} than does our proposal using the
null \jmin.

Spacelike surfaces are easily deformable into other spacelike
surfaces, whereas the same is not true for null surfaces (see
e.g.~\cite{bousso}).  Thus spacelike surfaces are in some sense less
constrained than null ones and thus maybe less appropriate in any
attempt, such as ours, to specify cosmological boundary conditions in
the most economical way.  In addition, the resulting universe would not obey
the Perfect Cosmological Principle, having a preferred time at
$t=0$. As discussed above, we also might conjecture that the quantum
state corresponding to null boundary conditions is simpler. In
general, however, there seems to be no strong argument against such a
boundary condition surface as compared to a null surface.

\section{Summary and conclusions}
\label{sec-conc}

	We have investigated the possibility of making
``future-eternal'' inflation eternal also to the past.  Starting with
a de Sitter spacetime background dominated by the false-vacuum energy
in which true-vacuum bubbles can form, we specify that the bubble
distribution at any time is in exactly the steady-state configuration
asymptotically approached by semi-eternal inflation. All bubbles are
then equivalent, and the statistical distribution of bubbles admits
both a time-translation and space-translation invariance in the
background inflating space.

The described region (``region I'') has no initial time, but has a
null boundary \jmin that is the limiting surface as $t\rightarrow
-\infty$ of the flat spatial sections. The steady-state configuration
at $t > -\infty$ corresponds to the inflaton field being in the false
vacuum state everywhere on \jmin, with no incoming radiation from
\jmin.  This surface can be reached by a past-directed geodesic of
finite proper length from any point in the space, so region I is
geodesically incomplete; this is the ``singularity'' pointed to by
theorems purporting that past-eternal inflation is
impossible~\cite{singth,bgv}.  But the space can (and should) be
extended, as the state on \jmin also constitutes boundary conditions for
the region past \jmin if the manifold is extended. These boundary
conditions imply that a duplicate copy of region I exists on the other
side of \jmin. Together region I and the new ``region II'' constitute
a geodesically complete (i.e. non-singular) inflating cosmology that
admits a coordinatization in which the background space and bubble
distribution are time-translation and space-translation invariant.

The null surface \jmin constitutes a cosmological boundary condition
surface on which the universe is classically in a particularly simple
and symmetric state.  Although \jmin is an infinite null surface, some
of its points are in causal contact, and one might attribute a finite
invariant volume of ${O}(H^{-3})$ to it.  Therefore specifying
boundary conditions on it is {\em not} like specifying them on an
infinite spacelike hypersurface (which would lead to a severe
``horizon problem'').

It might be possible to construct a quantum state corresponding to our
classical boundary conditions on \jmin by taking a null-limit of
spacelike sections on which the wave functional describing the fields
is centered on the desired classical state. But an explicitly null
quantum formulation of our null boundary proposal has not been
provided and would constitute an interesting future study.

It is widely thought that the ``arrow of time'' is connected with
cosmological boundary conditions.  In our model, time flows away from
\jmin, and the AOT is consistent among all physical observers that
can compare it. The AOT is not, however, defined globally, and in our
model the statistical description of the universe admits a global
symmetry that includes time-reversal.

This symmetry, along with the presence of two duplicate,
non-communicating universes, motivates---though does not
require---formally identifying antipodal points on the manifold.  We,
and others, have studied this identification on de Sitter space
classically and quantum-mechanically.  Antipodally identified (or
``elliptic'') dS has the virtues that it is causally stable and
observers have no event horizons.  It has the disadvantage that its
non-time-orientability makes defining a reasonable quantum field
theory difficult.  With antipodal identification our model is more
economical as the two duplicate universes are identified; however not
all of the attractive features of ``pure'' edS remain.

Our model can be generalized to other inflaton potentials (such as for
chaotic inflation), and therefore allows one to partially
understand the global structure of the eternally inflating spaces that
result.  One may also use an analogous construction to specify a
semi-eternally inflating but non-singular universe by placing boundary
conditions on a spacelike section.

Our primary conclusion is that it is possible, using only ``standard''
ingredients underlying popular models for inflation, to specify simple
cosmological boundary conditions on an infinite null surface that lead
to past- and future-eternal inflation.  Such a universe would obey the
same Perfect Cosmological Principle that governs a semi-eternally
inflating universe long after its beginning.  If our construction
survives scrutiny, and can be specified at the fully quantum (or
quantum-gravitational) level within a theory of fundamental physics,
it could serve as the basis for a realistic cosmology that avoids a
cosmological singularity, a beginning of time, or a creation of the
universe ``from nothing''.  

\medskip
\centerline{\bf Acknowledgments}
\medskip
We thank Alex Vilenkin for helpful comments on a draft of
this paper.  AA is supported by a grant from the W.M. Keck foundation.
SG is supported in part by US Department of Energy grant
DE-FG02-91ER40671.

\vskip0.3in
\appendix

{\bf APPENDIX}

Here we review Green functions and the initial value problem in some
detail, focusing for simplicity on the inhomogeneous scalar wave
equation in fixed background spacetimes.  We are interested in how
much information must be specified, and where, in order to fix the
field throughout spacetime.  

\subsection{Defining Green functions}

Imagine that our scalar field $\phi$ satisfies \ba \(\Box - m^2\) \phi
= q\(x^\mu\), \labeq{boxphi} \ea where $q$ is an arbitrary source
term, and we wish to determine $\phi$ at some point $P$ with
coordinates $x_P$.  This is possible given $\phi$ and its time
derivative on some complete spacelike slice through the past
light-cone (PLC) of $P$.  Let us see why this is so, and whether
this is the only suitable set of initial data for the problem.

First introduce another function $G$, which depends on $P$ and is
assumed to satisfy
\begin{equation}
\(\Box-m^2\) G = s\(x^\mu\),
\labeq{boxG}
\end{equation}
where $s$ is to be chosen.
Taking $G$ times~\eqn{boxphi} minus $\phi$ times~\eqn{boxG} and
integrating through some four dimensional volume ${\cal V}$ yields
\begin{eqnarray}
\int d^4 x \partial_\mu\( G \sqrt{-g} g^{\mu \nu} \partial_\nu \phi -
\phi \sqrt{-g} g^{\mu \nu}  \partial_\nu G \) \nonumber \\
= \int \sqrt{-g} \( q G -
s \phi \) d^4 x.
\labeq{intform}
\end{eqnarray}
This allows determination of
$\phi(P)$ in terms of its values elsewhere by a suitable choice of
$s$, $G$, and ${\cal V}$.

To effect this, we must isolate $\phi(P)$, either on the left or right
hand side of~\eqn{intform}.  Let us first use the RHS. Take $s$ to
be a $\delta$-function at $P$ (or, more carefully, a function peaked
near $P$ that can be taken to a $\delta$-function limit).  Then
choosing ${\cal V}$ to enclose $P$ gives $\phi(P)$ for the second term
on the RHS of Eq.~\eqn{intform}. Using Gauss's law, this equation
can now be used to express $\phi(P)$ as a surface integral over of
$\phi$ and its derivative over the boundary $\partial{\cal V}$ of
${\cal V}$, plus a volume integral over ${\cal V}$ of the source
alone.

To fix $G(x)$, we must choose boundary conditions specific to our
choice of $s$. Two conventional choices are to require $G$ to vanish
either to the future or to the past of $P$.  In the first case
$G$ is known as the ``retarded'' Green function and may be denoted
$G_R$; in the second case it is ``advanced'' ($G_A$).

\begin{figure}
\includegraphics[width=8.5cm]{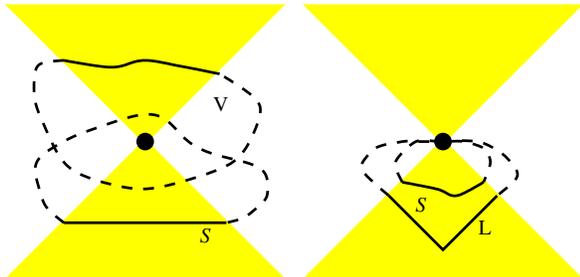}
\caption{Possible choices of ${\cal V}$ for
different Green functions.  The solid lines indicate the regions
${\cal S}$ of $\d {\cal V}$ which contribute to the integral.  On the
left, we have volumes suitable for the retarded and advanced Green
functions $G_R$ and $G_A$, and on the right we have volumes suitable
for the commutator function $G_C$.}
\label{fig-green}
\end{figure}

We must now choose ${\cal V}$ (see Fig.~\ref{fig-green}).  It turns
out (see, e.g., ~\cite{courhil}) that $G_A\neq 0$ only on and within
the FLC of $P$, and that $G_R\neq 0$ only on and within the PLC of
$P$. Hence only a segment ${\cal S}$ of $\partial {\cal V}$
contributes to Eq.~\eqn{intform}, enabling us to deduce $\phi(P)$
using only data on some connected surface making a complete span
through either the FLC or the PLC of $P$. The usual choice, consistent
with our standard ideas about causality, is to pick a spacelike
surface of constant time (see left side of Fig.~\ref{fig-green}).
Then evaluation of Eq.~\eqn{intform} requires $\phi$ and $\dot{\phi}$
on ${\cal S}$. For a more general $\partial V$, we require $\phi$ and
its normal derivative on ${\cal S}$.  (Note that this surface need not
be everywhere spacelike, but if not then the data must be
self-consistent.)  The case of present interest is that of a null
surface, for which the normal lies everywhere within the surface
itself (this is made possible by the Lorentzian signature of the
spacetime.)  Then specifying $\phi$ on the surface also specifies
the normal derivative of $\phi$ on it.  A suitable null surface is, for
example, the forward light-cone $L$ of a point $Q$ to the past of $P$;
now we need only know $\phi$ on the surface ${\cal S}$ where this FLC
of $Q$ is within the PLC of $P$ (see right side of
Fig.~\ref{fig-green}). (If the data near $Q$ is suitably regular, the
non-smooth nature of the surface at $Q$ is unimportant.)  In the same
spirit, one might consider a ``wedge'' such as the boundary of the
future of the segment of a line in the past of $P$.
   
Let us now consider the second way of isolating $\phi(P)$, using the
LHS of Eq.~\eqn{intform}.  Set $s=0$, and choose some piece of
$\partial {\cal V}$ to be spacelike and run through $P$.  Then choose
$G$ to vanish everywhere on this surface. Further, choose it to have a
$\delta$-function at $P$ in its normal derivative; the integral then
picks up a contribution proportional to $\phi(P)$.  This choice
comprises boundary conditions for $G$, and the function they
determine, which we denote $G_C$, now vanishes outside the (full)
light-cone of $P$ and satisfies the homogeneous version of
Eq.~\eqn{boxG}.  Choosing now the remainder of $\partial V$ to close
off to the future or to the past of $P$, we again may determine
$\phi(P)$ once data is given on a complete slice ${\cal S}$ through
one of the light-cones.  $G_C$ may be called the commutator function,
because it turns out to be equal to $-i$ times the commutator of a
free quantum field (at least in a globally hyperbolic spacetime).
Note also that $G_C=G_A-G_R$: subtracting the (inhomogeneous)
equations governing $G_R$ and $G_A$ leaves a homogeneous equation for
the difference.  Then the difference must vanish outside the
light-cone of $P$, just as for $G_C$.

In all cases, $G$ depends on the point $P$, and so may
be considered as a function of two variables, $x$ and $x_P$.  Because $\phi$
satisfies~\eqn{boxphi}, it turns out that the $G$ also
satisfies~\eqn{boxphi} with respect to $x_P$, at least away from
$x_P=x$.

\subsection{Green functions for Minkowski and dS spacetimes}

Let us now outline a procedure to obtain some Green functions for
massive scalar fields in certain spacetimes, using a slightly
unconventional but perhaps more intuitive method not relying on the
usual Fourier techniques or $i\epsilon$-prescriptions.  We start with
the commutator function for a massless field in 3+1 D Minkowski space,
\begin{equation}
 G_C^{M_0}=\sgn \(t\) \delta \( t^2-r^2\) / 2\pi,
\labeq{gfmink}
\end{equation}
with coordinates such that $P$ is at the origin. That this solves
$\Box \phi=0$ can be seen by expanding out the $\delta$-function and
comparing to a general superposition of incoming and outgoing
spherically symmetric waves.  (A nice discussion of this function is
found in~\cite{diracqft}.)  Note that this is only non-zero on the
light-cone itself, vanishing even inside the cone.  This is a special
property of massless fields in even-dimensional spacetimes
(see~\cite{courhil} for more details.)

To obtain the solution for the massive field in Minkowski space, we
start by considering
$$G_1 \equiv f\(t^2-r^2\) \sgn\(t\) \Theta \(t^2-r^2\). $$
The $\sgn$ and $\Theta$ functions enforce the gross features of the behaviour
that we require. Writing $\Box$ using spherical polar coordinates, we 
find \begin{eqnarray}
(\Box-m^2)G_1&=&\sgn \(t \) \Theta \(t^2-r^2\) \(\Box-m^2\) f \nonumber \\
&-& 4 f\(0\) \sgn(t) \delta\(t^2-r^2\).
\labeq{gross}
\end{eqnarray}
Let us choose $f(\tau^2)$ (where here $\tau^2=t^2-r^2$) such that
$\(\Box-m^2\) f=0$ inside the light-cones; a general solution is
$f(\tau^2)=AJ_1(m\tau)/\tau+BY_1(m\tau)/\tau$ where $J_1$ and $Y_1$
are Bessel functions as in Ref.~\cite{abram}. The first term of
Eq.~\eqn{gross} is then zero.  For the second term, we notice that it
is basically the solution to the massless problem that we've already
found.  So let us simply add the massless solution $G_C^{M_0}$ from
Eqn.~\eqn{gfmink} to our ansatz $G_1$.  Since $\Box G_C^{M_0}=0$, we
have left $-m^2G_C^{M_0}$ on the RHS.  By choosing $A=-m/4\pi$ and
$B=0$, we obtain a complete cancellation,
leaving us with our desired solution. Thus the full Green function is
\begin{equation}
G_C^M={\sgn \(t\) \over 2\pi}\( \delta \(\tau^2\)-{m\Theta(\tau^2)J_1(m\tau)\over2\tau}\).
\labeq{gfminkm}
\end{equation}

We may now consider generalizing Eq.~\eqn{gfminkm} to other spacetimes
such as dS. First, let us note that we took $f$ above to be a function
of the proper time $\tau$ from the origin, within the light-cone.
Thus the commutator function was invariant under Lorentz boosts.  This
suggests that in dS, for example, we make our commutator inside the
forward light-cone invariant under isometries which leave $P$ fixed.
This is the same as saying that it should be a function of the proper
time from the origin alone.  One can define this quantity in terms of
an ``angle'' in the 5D embedding space~\cite{dsreview} (see footnote to
Sec~\ref{sec-horiz}), but perhaps a more intuitive way to proceed is
as follows.  A patch of dS can be covered by coordinates in which the
metric is the same as that for an open expanding FRW universe, with
scale factor $H^{-1}\sinh Ht$.  The proper time along the geodesics
representing comoving observers is just given by the coordinate time
$t$.  These geodesics
all intersect at a point as $t\rar 0$, which we choose to be $P$, and
they cover all of the interior of the FLC of $P$.  We therefore need only
find an appropriate spatially homogeneous solution of the massive wave
equation in the coordinate patch of the open slicing of dS.  The
equation reads:
\begin{equation}
\frac{1}{\sinh^3 Ht } \d_t \( \sinh^3 Ht \d_t \phi \)+ m^2 \phi=0.
\end{equation}       
To solve this~\cite{eucmode}, write $\phi= H\chi / \sinh Ht$ and set
$z=\cosh Ht$ to obtain
\begin{equation}
(1-z^2) \chi_{,zz}- 2 z \chi_{,z} + \(2-{m^2\over H^2} - \frac{1}{1-z^2}\)
\chi=0,
\end{equation}
which is Legendre's equation~\cite{abram} with $\nu= -1/2 + \sqrt{9/4
- m^2/H^2}$ and $\mu = -1$. The solution to this equation which is
regular as $z\rar 1$ is $P^{\mu}_{\nu} (z)$.  In terms of $t$, this
tends to $Ht/2$ as $t\rar 0 $ independent of $m^2$, hence $\phi$ tends
to $H/2$.

We can now use this result to deduce the form of the commutator over
all of dS space.  Extend the meaning of $\tau^2$ to be the
signed geodesic distance squared between $P$ and the point in
question.  Note that we do not consider the point to be in the
light-cone of the antipode of $P$, since no geodesic exists which
connects it to $P$.  In this case we rather define $G_C$ to be zero
(but see our discussion of the antipodal quantum commutator above.)
Near $P$ the space is locally Minkowskian, so we can compare to our
Minkowksi result to obtain:

\begin{equation}    
G_C^{dS}={\sgn(t)\over 2\pi}\[\delta(\tau^2)-
\frac{m^2\Theta(H^2\tau^2)}{2\sinh H\tau} P^{-1}_\nu \(\cosh H\tau\)\].
\labeq{gfdsfull}
\end{equation}

Here by $t$ is meant some suitable generalization of Minkowski time,
with $\sgn(t)$ thus serving to make $G_C^{dS}$ be of opposite sign in
the future and past light-cones, and $\tau=+\sqrt{\tau^2}$. The
$\sgn(t) \delta(\tau^2)$ should be interpreted as the generalization
of the equivalent term in the Minkowski result.

\subsection{Domains of dependence}

Given our Green functions, we would like to investigate to what extent
fixing the fields on a null-cone $L$ of a point $Q$ determines the
field values within that cone.  We are particularly interested in the
importance of the field near $Q$, as our cosmological boundary surface
\jmin can be considered the light-cone with $Q$ ``at'' past-timelike
infinity.

Take, as a first example, the Green function $G_R$ for a massless
field such as the scalar field in 4D Minkowski space with $G_R$ from
Eq.~\eqn{gfmink}. Because $G_R$ has support only on the PLC of $P$,
the field at $P$ depends only on the intersection of $P$'s PLC and
$L$. This indicates two things. First, specifying data on $L$
explicitly determines the field everywhere inside $L$ (i.e. everywhere
in the future of $Q$). Second, the field at any point $P$ inside $L$
does not depend on $\phi(Q)$; we might thus consider $Q$ to be
irrelevant in terms of what occurs within $L$.  This, however, hides a
subtlety: while $G_r$ and $\phi$ on $L$ always allow the construction
of a valid solution of the field equations, there is no guarantee that
the field so obtained is continuous with the field specified on $L$
unless some assumptions of field regularity at the vertex ($Q$) are
made: if we wish to evaluate $\phi$ {\em on} $L$ to check that we have
a genuine solution to our boundary value problem, then we must either
know $\phi(Q)$, or assume that the field is regular as $Q$ is
approached along $L$, so that we can extend the field to $Q$ by
continuity.

For a massive field the strict ``Huygen's Principle'' does not hold;
while much of the contribution to the integral in Eq.\eqn{intform}
comes from the PLC of $P$ where $G_R$ is singular
(see~\cite{courhil} for some discussion of this ``generalized Huygen's
principle''), there is also a contribution from inside $P$'s
PLC because $G_R$ is everywhere nonzero there. 
	
In the dS case, $G_R$ as given by Eq.~\eqn{gfdsfull} falls off
exponentially as $\tau\rightarrow \infty$; thus we may expect that
when $L$ is taken to be \jmin, where the vertex lies ``at''
$\tau=\infty$, the field values inside \jmin will not depend on the
field at the vertex (where by ``at the vertex'' we mean within \jmin
in the limit $\tau\rightarrow \infty$).  More explicitly, we may
consider a boundary surface comprised of \jmin at $\tilde t > \tilde
t_0$ for some $\tilde t_0$, closed by the spacelike surface $\tilde
t=\tilde t_0$, where $\tilde t$ and $\tilde t_0$ are in the global
closed coordinates.  Then the field integral~\eqn{intform} consists of an
integral along \jmin, and an integral over the $\tilde t=\tilde t_0$
hypersurface, which has a finite physical volume of order $H^{-3}$.
But then as $\tilde t_0 \rightarrow -\infty$, $\tau \rightarrow
\infty$, so this second contribution will vanish unless either the
average of the field or its $\tilde t$-derivative blow up faster than
$G_R^{-1}$.  Hence we expect that the field at any point in the space
will depend {\em only} on the values on \jmin, and not on the field
behavior at past infinity, as long as the fields are assumed to be
suitably finite and regular.

This argument might also be applicable to gravity; the equations
governing the Weyl conformal tensor can be cast, at the linear level,
as wave-equations for a spin-2 field, and boundary conditions can be
specified on a null surface such as \jmin~\cite{pr}. We therefore
expect that our boundary conditions on \jmin determine the Weyl tensor
uniquely, and therefore determine the spacetime to be pure dS when
bubble nucleations are neglected.

\end{document}